\documentclass[10t]{JHEP3}
\usepackage{amssymb,amsmath}
\usepackage[]{graphicx}
\usepackage{epsfig}
\usepackage{amsfonts}

\usepackage{amscd}

\def\be{\begin{eqnarray}}
\def\ee{\end{eqnarray}}

\def\half{\frac{1}{2}}
\def\d{\partial}

\newcommand{\idn}{{1\relax{\kern-.35em}1}}

\DeclareMathOperator{\tr}{Tr}

\def\tq{\tilde q}
\def\tc{\tilde \chi}
\def\tr{\tilde \rho}
\def\tX{\tilde X}
\def\tY{\tilde Y}
\def\ts{\tilde \sigma}
\def\d{\delta}

\preprint{WIS/08/08-APR-DPP }

\title{On SQCD with massive and massless flavors}
\author{Amit  Giveon,$^{1}$ Andrey Katz$^{2}$ and Zohar Komargodski$^{3}$\\
$^{1}$ Racah Institute of Physics, The Hebrew University, Jerusalem 91904, Israel\\
$^{2}$ Physics Department, Technion-Israel Institute of Technology, Haifa 32000, Israel\\
$^{3}$ Department of Particle Physics, The Weizmann Institute of Science,
Rehovot 76100, Israel\\
E-mails: \email{giveon@phys.huji.ac.il},\ \email{andrey@physics.technion.ac.il},\
\email{zkomargo@wisemail.weizmann.ac.il}}
\date{\today}

\abstract{We consider supersymmetric QCD in the free magnetic phase with massless and massive flavors. The
theory has a supersymmetry breaking pseudo-moduli space of vacua and a runaway behavior far away from the
origin. A two-loop computation reveals that the origin is destabilized and there is no meta-stable SUSY breaking
solution. We also study the embedding of this model in type IIA string theory and find evidence for similar
behavior. The perturbative brane dynamics involves simple interactions between branes, correctly predicting the
two-loop result in the gauge theory. Our results also apply to the case when all the flavors are massive but
have hierarchy among them, leading to possible instability which is manifest both in field theory and the brane
description. }


\begin{document}

\maketitle

\section{Introduction}\label{intro}

Dynamical SUSY breaking (DSB) may be an attractive explanation of the hierarchy between the electro-weak scale
and the Planck scale~\cite{Witten:1981nf}. In spite of the fact that there are models which break SUSY
spontaneously at the ground state (see for example the review~\cite{Shadmi:1999jy}), these models are extremely
rare and they are subject to some severe constrains. Moreover, a generic (and calculable) SUSY breaking model
should have a spontaneously broken R-symmetry~\cite{Nelson:1993nf}, leading to a relatively light R-axion, which
gets its mass only from couplings to supergravity~\cite{Bagger:1994hh} or to some higher dimensional operators.
Current astrophysical observations disfavor this possibility.

These difficulties provide a good motivation to consider the possibility that SUSY is broken dynamically in a
\emph{meta-stable} vacuum (see~\cite{Intriligator:2007cp} for a review of DSB both in stable and meta-stable
vacua). The idea of DSB in a meta-stable vacuum got a lot of attention after it was shown by Intriligator,
Seiberg and Shih (ISS) in~\cite{Intriligator:2006dd} that a simple and generic class of models, supersymmetric
QCD (SQCD) in the free magnetic range, possesses local meta-stable SUSY breaking vacua (see
e.g.~\cite{Intriligator:2007py,mass splittings,pheno,Giveon:2007ef}). The authors of~\cite{Intriligator:2006dd}
considered SQCD with gauge group $SU(N_c)$ and $N_f$ flavors in the range $N_c<N_f<3N_c/2$. It was shown that if
the quarks are given small and \emph{equal} masses then the theory has a long-lived SUSY breaking vacuum. Since
in this range the magnetic description of the theory is weakly coupled in the IR, the analysis at low energies
was done using the Seiberg duality~\cite{Seiberg:1994pq}. At tree-level massive SQCD has a moduli space of SUSY
breaking vacua, but one-loop quantum effects stabilize the pseudo-moduli at the origin of field space.

In this work we study SQCD with \emph{massless} and massive (but light) flavors. There are several reasons to
consider this model. First, such a theory is an extreme case of massive SQCD with generically distributed
masses. Even though the ISS model looks as a good starting point for direct gauge mediation (for early models
see~\cite{Affleck:1984xz,Poppitz:1996fw,Dimopoulos:1997ww,ArkaniHamed:1997fq}), it should be modified in order
to produce viable phenomenology. One of the reasons for such a modification is an approximate R-symmetry at low
energies which protects gauginos from getting masses. Some of the modifications included hierarchical quark
masses~\cite{mass splittings} since this introduces another scale with which one can tune the lifetime
independently of gaugino masses. Therefore, it is interesting to understand a limiting case of a generic
distribution, taking some of the flavors to be completely massless.

Another motivation is that SQCD has a natural embedding in type IIA string theory, as the low energy theory on
intersecting Neveu-Schwarz (NS) fivebranes and D-branes~\cite{Elitzur:1997fh} (for a review,
see~\cite{Giveon:1998sr}). The brane description of the meta-stable vacua of~\cite{Intriligator:2006dd} was
studied recently in~\cite{Ooguri:2006bg,Franco:2006ht,Bena:2006rg,Giveon:2007fk}. Many modifications and
variations of the basic model of~\cite{Intriligator:2006dd} were constructed, along with their brane
descriptions (see e.g. \cite{Franco:2006es,branes,Giveon:2007ew}). It is tantalizing that in these examples one
could identify identical patterns of meta-stable SUSY breaking states in the gauge theory and the classical
brane system. In gauge theory, pseudo-moduli are stabilized by one-loop quantum
effects~\cite{Intriligator:2006dd,Intriligator:2007py,Shih:2007av}, while in the classical brane dynamics
regime, gravitational attraction in the NS fivebrane background stabilizes the branes in long-lived SUSY
breaking meta-stable configurations~\cite{Giveon:2007fk,Giveon:2007ew}. It is interesting to see whether this
correspondence can be pushed further, and to check whether also our system has similar qualitative properties in
the perturbative brane dynamics regime.

We consider $SU(N_c)$ SQCD with $N_{f0}$ massless flavors and $N_{f}-N_{f0}$ massive ones in the range
$0<N_{f0}<N_c<N_f<3N_c/2$ and study it in the dual magnetic description. In such a case the maximal possible
rank of the quarks mass matrix is still larger than the rank of the dual gauge group, hence, there is no
classical supersymmetric solution. Instead, classically, these models possess a moduli space of SUSY breaking
vacua. However, the pseudo-moduli associated with the massless (electric) quarks are {\it not} lifted by
one-loop quantum effects in field theory~\cite{Franco:2006es} and a two-loop calculation is required to decide
what is the fate of this system.

In this paper we perform the calculation of the two-loop effective potential for these pseudo-moduli. We show
that at the two-loop level these directions are destabilized and, consequently, there is no SUSY breaking
meta-stable vacuum near the origin. This result is also important for the case when all the flavors are massive,
but there is mass hierarchy among them. In that case the two-loop contribution of heavy quarks will dominate the
one-loop contribution of light quarks and the SUSY breaking solution of~\cite{Intriligator:2006dd} may be
destabilized.

In addition, we study the corresponding brane description and find compelling evidence that a similar
instability occurs there. In particular, in the appropriate sense, the ``origin" is destabilized  by the brane
dynamics. Note that so far in all the studied examples it was found that there is a non-trivial correspondence
between the weakly coupled brane dynamics and field theory: whenever there is a meta-stable state in gauge
theory one could identify a meta-stable state in the classical branes picture. Our work provides another
non-trivial check of this correspondence, beyond one-loop effects in field theory (and beyond classical gravity
in the brane dynamics). We emphasize that understanding the perturbative brane dynamics involves simple
classical considerations, correctly predicting the result of an intricate two-loop evaluation in field theory.

This paper is organized as follows. In section~\ref{field theory} we study a simple Wess-Zumino (WZ) model which
has a similar structure to the low energy SQCD, and then turn on the appropriate gauge interactions. In
section~\ref{braneembedding} we present (after a brief review) the brane description of this gauge theory and
analyze it. In section~\ref{comments} we comment on some implications of our results to a theory with a general
distribution of masses, especially the issue of stability of the local minimum of massive SQCD and its brane
construction. Finally, we summarize in section~\ref{summary}. Appendix~\ref{append} contains a brief review
of~\cite{Martin:2001vx} and some technical details related to our calculation.

%%%%%%%%%%%%%%%%%%%%%%%%%%%%%%%%%%%%%%%%%%%%%%%%%%%%%%%%%%%%%%%%%%%%%%%%%%%%%%%%%%%%%%%%%%%%%%%%%%%%%%%%%%%
\section{Field Theory Analysis}\label{field theory}
\subsection{A Simplified Wess-Zumino Model}\label{WZsec}

In this section we analyze a simple WZ model, which has a pseudo-moduli space of SUSY breaking vacua at the
one-loop approximation, and show that it \emph{does not} have any SUSY breaking minimum near the origin of field
space. Consider a model with the chiral superfields listed in Table \ref{charges}, and a superpotential
\begin{equation}
   \mathcal W=hq^i\Phi_i^j\tilde q_j-h\mu^2(\Phi_{11}+\Phi_{22}),
\end{equation}
where $i,j=1,2,3$.
The components of the matrix $\Phi$ and vectors $q,\tq$ are given in terms of the fields in Table \ref{charges} by
\begin{equation}
\Phi=\begin{pmatrix}\Phi_{11} & \Phi_{12} & X \\ \Phi_{21} & \Phi_{22} & Y
\\ \tX & \tY & Z \end{pmatrix},\hspace{1 em} q=\begin{pmatrix}\chi & \rho & \sigma\end{pmatrix},
\hspace{1 em}
\tq=\begin{pmatrix}\tc \\ \tr \\ \ts\end{pmatrix}.
\end{equation}
Let the K\"ahler potential be canonical.
The parameter $h$ controls our loop expansion. Of course, the physical
parameter corresponding to $h$ is IR free, allowing a faithful perturbative treatment.

The model has manifest $SU(2)\times U(1)_{\chi\rho}\times U(1)_\sigma\times U(1)_{\ts} \times U(1)_R$ symmetry,
where the two upper components of $q$ and $\tilde q$ transform as fundamentals of the $SU(2)$ and $\Phi$
transforms in the adjoint.\footnote{By that we mean that the upper-left $2\times 2$ submatrix of $\Phi$ sits in
the adjoint, $(X,Y)$ and $(\tX,\tY)$ are fundamentals of $SU(2)$ and $Z$ is neutral.} Under the various $U(1)$
symmetries the fields transform as summarized in Table \ref{charges}. Note that once $\mu$ is turned off there
is an $SU(3)^2\times U(1)_B\times U(1)_R$ symmetry. The baryon number is still present in our model as
$U(1)_{\chi\rho}+U(1)_{\ts}-U(1)_{\sigma}=U(1)_B$.

\begin{table}\begin{center}\begin{tabular}{c|cccc} & $U(1)_{\chi\rho}$ & $U(1)_\sigma$ & $U(1)_{\ts}$ & $U(1)_R$\\ \hline
$\Phi_{11}$, $\Phi_{12}$, $\Phi_{21}$, $\Phi_{22}$ &0  &0  & 0& 2\\
$X$,$Y$&1 &0 &-1 & 2\\
$\tX$, $\tY$&-1&-1 &0 &2 \\
$\chi$, $\rho$& -1& 0&0 & 0\\
$\tc$, $\tr$& 1& 0& 0& 0\\
$\sigma$&0 &1 &0 &0\\
$\ts$&0 &0 &1 &0\\
$Z$ &0 &-1 &-1 &2
\end{tabular}\caption{The chiral superfields and their global $U(1)$ charges.}\label{charges}\end{center}\end{table}

This system has no classical SUSY preserving vacuum. The F-terms of the relevant meson components are
\begin{equation}
    \frac{\partial \mathcal W}{\partial\Phi_{ij}}=h\begin{pmatrix}\tc \\
\tr\end{pmatrix}\begin{pmatrix}\chi & \rho \end{pmatrix} -h\mu^2\mathbb{I}_{2\times 2},\hspace{2em}
i,j=1,2.\end{equation} The first term is at most of rank one while the second term is of rank two.
Thus, SUSY
is broken by a rank condition. Nonetheless, there is a stationary point with positive energy (i.e. spontaneously broken SUSY),
\begin{equation}\label{classical solution}
    \Phi=0, \hspace{1em} q=\begin{pmatrix}\mu & 0 & 0\end{pmatrix},\hspace{1em} \tq=\begin{pmatrix}\mu \\ 0 \\
    0\end{pmatrix}.
\end{equation}
The global symmetry is broken as $SU(2)\times U(1)_{\chi\rho}\hookrightarrow U(1)'$. Thus, the above classical
solution has three Goldstone bosons. In addition, classically, there are some dangerous pseudo-flat directions.
Our purpose is to understand their quantum mechanical fate. {}For convenience, we take $\mu$ and $h$ to be real and define
 \begin{equation}\rho_{\pm}=\frac1{\sqrt2}(\rho\pm\tilde \rho), \hspace{1.5em} \chi_\pm=\frac1{\sqrt2}(\chi\pm\tilde
 \chi).
 \end{equation}
The squared mass of the fields $\sigma,\ts,X,\tX,\Phi_{12},\Phi_{21}$ is
$h^2\mu^2$. Similarly, the mass of $\chi_+,\Phi_{11},$ $\Im\rho_+,\Re\rho_-$ is $2h^2\mu^2$.
The three real Goldstone bosons are $\Im\chi_-,\Re\rho_+,\Im\rho_-$. All the rest are ``accidental''
pseudo-moduli which are not protected quantum mechanically, in general.

The results of the one-loop effective potential in this model are known from \cite{Franco:2006es} and we shall
review them here. All pseudo-moduli fields but $Z$ obtain similar positive mass squared terms,
$$m^2_{\Re\chi_-}=m^2_{\Phi_{22}}=2m^2_{Y,\tY}=h^4\mu^2\,\frac{\ln4-1}{8\pi^2}.$$
However, $Z$ remains massless at one-loop. One can argue that in the one-loop effective potential there
will be no $Z^n$ terms, for any $n>1$. The way to see it is to turn off the expectation value of the classical
pseudo-moduli $Y$ and $\tY$. Doing so, particles whose mass depends on the expectation value of $Z$ are
decoupled (in the mass matrix) from the $\rho$ sector which breaks SUSY. Thus, they sit in supersymmetric
multiplets and the one-loop contribution vanishes identically.

This means that in order to understand the dynamics of this model it is necessary to compute the two-loop
effective potential along the pseudo-moduli space parameterized by $Z$. Explicitly, we replace all the fields by
their fluctuations and assume, without loss of generality, that the field $Z$ obtains a real expectation value
around which it fluctuates. Indeed, we can use the symmetry generator $U(1)_\sigma+U(1)_{\ts}$ to rotate the
point where $Z$ is real to any other complex value of $Z$ with the same magnitude (Note that all the other
expectation values vanish since they correspond to fields which are massive at tree-level or one-loop.).

The superpotential is given by
\begin{eqnarray}\label{fullsuperpotential}
\mathcal{W}&=&-h\mu^2\d\Phi_{22}-h\mu^2\d\Phi_{11}+\\ \nonumber
&+&\frac h2\begin{pmatrix}\d\chi_++\d\chi_- +\sqrt2\mu \\
\d\rho_++\d\rho_- \\ \sqrt2\d\sigma\end{pmatrix}^T \begin{pmatrix}\d\Phi_{11} & \d\Phi_{12} & \d X \\
\d\Phi_{21} & \d\Phi_{22} & \d Y
\\ \d\tX & \d\tY & Z+\d Z \end{pmatrix} \begin{pmatrix}\d\chi_+-\d\chi_-+\sqrt2\mu\\ \d\rho_+-\d\rho_-\\
\sqrt2\d\ts\end{pmatrix}.
\end{eqnarray}
 The spectrum of masses is as quoted above (when expanding around $Z=0$), except that the fields
$\sigma,\ts,X,\tX$ mix in a simple manner. The mass eigenstates are given by some linear combinations
\begin{equation}
\d\sigma=\sin\theta\ \d A+\cos\theta\ \d B,\hspace{2em}\d X=\cos\theta\ \d A-\sin\theta\ \d B,
\end{equation}
and analogous equations for the tilded fields (with the same mixing angles). Hereafter we use the notations
$s_\theta\equiv\sin\theta$ and $c_\theta\equiv\cos\theta$.  $A$ and $B$ are mass eigenstates with the following
masses
\begin{equation}\label{masses}
m^2_{A,B}(Z)=h^2\left(\mu^2+\frac{Z^2}2\mp \frac Z2\sqrt{Z^2+4\mu^2}\right).
\end{equation}
The mixing angle is
\begin{equation}
s_\theta^2=\frac{h^2\mu^2-m_A^2}{m_B^2-m_A^2}.
\end{equation} The mass spectrum of all the particles except
$\rho_{\pm}$ is supersymmetric.

{}From now on, the two-loop evaluation is, in principle, straightforward (but in practice there are many
diagrams). All the required two-loop functions and diagrams are beautifully described in \cite{Martin:2001vx};
some highlights are reviewed in Appendix A. We have simplified the computational task (in particular, the number
of diagrams) with a few tricks which may be useful also in other models.

\begin{figure}[htbp]
\begin{center}
\epsfig{file=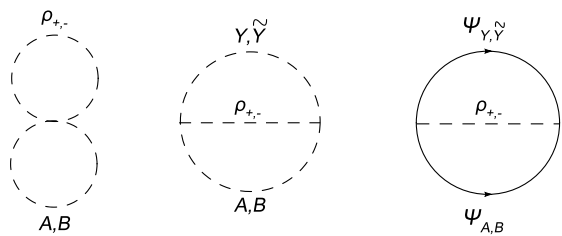,scale=0.7} \caption{\label{fig1} The only $3$ two-loop diagrams contributing to the $Z$
dependent part of the effective potential. Conforming with \cite{Martin:2001vx}, we refer to them as SS, SSS and
FFS, respectively.}
\end{center}
\end{figure}

Consider a different theory in which we switch off the linear term for $\Phi_{22}$ in the superpotential
$\mathcal W$. In other words, we consider a theory whose superpotential is $\mathcal W'=\mathcal
W+h\mu^2\d\Phi_{22}$, where $\mathcal{W}$ is given by (\ref{fullsuperpotential}). In this model the moduli space
parameterized by $Z$ still exists but now it is a supersymmetric moduli space. It cannot be lifted by
perturbative quantum corrections. This means  that the effective two-loop potential vanishes identically as a
function of $Z$. Thus, we can write the trivial equation,
\begin{equation}\label{twolooptrick}
    V^{(2)}_{\mathcal W}=V^{(2)}_{\mathcal W}-V^{(2)}_{\mathcal W'}~,
\end{equation}
for the two-loop effective potential we are after, $V^{(2)}_{\mathcal W}$. Note that all of the Yukawa, cubic
and quartic interactions are identical in the two models.
In fact, the only difference is that the fields $\rho_\pm$
of the model $\mathcal W'$ are in supersymmetric multiplets with mass $h^2\mu^2$. Consequently, diagrams that do
not cancel on the right hand side of (\ref{twolooptrick}) contain necessarily a $\rho_\pm$ {\it scalar}.
However, this is not the only simplification we can make. Since we want the diagrams to have some $Z$
dependence, we should better have an $A$ or $B$ (fermion or boson) running in the loop. Otherwise, the diagram
contributes only to the overall zero-point energy which we are not interested in. In this way we remain with
only three different diagrams! They are depicted in Fig.~\ref{fig1}.

Note that in general there is another possible topology for a two-loop diagram, one that includes mass flips for
fermions (the third diagram depicted in Fig.~\ref{4dg}). Even though the fermionic mass term in our theory is
not diagonal this diagram is absent because there is always a $\psi_Y$ fermion in the loop, which is massless.

At this stage, it remains to evaluate the coefficients of the diagrams in Fig.~\ref{fig1}. Of course, as follows
from (\ref{twolooptrick}), we must subtract from each of the diagrams the corresponding diagram in the theory
$\mathcal W'$. In terms of the functions given in~\cite{Martin:2001vx} and reviewed in Appendix~A, we get
\begin{equation}
V^{(2)}=V^{(2)}_{SS}+V^{(2)}_{SSS}+V^{(2)}_{FFS},
\end{equation}
where
\begin{gather}\label{SS}
V^{(2)}_{SS}= h^2s_{\theta}^2
    \left(f_{SS}(2h^2\mu^2,m^2_A)-2f_{SS}(h^2\mu^2,m^2_A)\right)+(A,s^2_\theta)\leftrightarrow (B,c^2_\theta),
\end{gather}
\begin{gather}\label{SSS}
V^{(2)}_{SSS}=h^4(\mu
c_\theta+Zs_\theta)^2\bigl(f_{SSS}(0,0,m^2_A)+f_{SSS}(0,2h^2\mu^2,m_A^2)-2f_{SSS}(0,h^2\mu^2,m_A^2)\bigr)+
\cr+(A,c_\theta,s_\theta)\leftrightarrow(B,-s_\theta,c_\theta),
\end{gather}
\begin{gather}\label{FFS}
    V^{(2)}_{FFS}=h^2s^2_\theta\bigl(f_{FFS}(0,m_A^2,0)+f_{FFS}(0,m_A^2,2h^2\mu^2)
    -2f_{FFS}(0,m_A^2,h^2\mu^2)\bigr)+\cr+(A,s_\theta^2)\leftrightarrow (B,c^2_\theta).
\end{gather}

Our results manifestly look like the difference of amplitudes in two different theories.
{}From here on, it is
straightforward to expand these functions in a Taylor series and to get that the overall contribution to the effective potential is (promoting $Z$ to be complex again)
\begin{equation}\label{vtwo}
V^{(2)}={\rm const}+h^6\mu^2\left(-1-\frac{\pi^2}{6}+\ln4\right)|Z|^2+{\cal O}(|Z|^4).
\end{equation}
Thus, the origin is destabilized. An examination of the effective potential as a function of $Z$ shows that
there is no minimum around the origin; the effective potential decreases monotonically.

Not surprisingly, there is no dependence on the renormalization scale, $Q$, in front of $|Z|^2$. It is a
consequence of the following RGE argument. The effective potential satisfies an equation of the schematic form
$$\left(Q\frac{\partial}{\partial
Q}+\beta_h\frac{\partial}{\partial h}-\gamma_\phi\phi\frac{\partial}{\partial \phi}\right)V=0,$$ where $\beta_h$
is the beta-function of a (physical) coupling $h$ and $\gamma_\phi$ is the anomalous dimension of a field
$\phi$. Since both of these functions begin at one-loop order (or higher), and since there is no $|Z|^2$ term at
one-loop or at tree-level, $Q\frac{\partial}{\partial Q}$ should annihilate the two-loop coefficient of $|Z|^2$
(or higher powers of $Z$), as indeed happens. Such an RGE argument is more general: loosely speaking, this means
that whenever a physical effect appears for the first time in the effective potential, it must be
renormalization scheme independent.

%%%%%%%%%%%%%%%%%%%%%%%%%%%%%%%%%%%%%%%%%%%%%%%%%%%%%%%%%%%%%%%%%%%%%%%%%%%%%%%%%%%%%%%%%%%%%%%%%%%%%%%%%%%%

\subsection{Supersymmetric QCD}\label{full model}
Our general model is SQCD, whose UV electric description is given by the superpotential
\begin{equation}\mathcal{W}=\sum_{a=1}^{N_f}m_{(a)}Q_a\tilde Q^a~,\end{equation}
where $Q_{a}$ ($\tilde Q_{a}$) is in the (anti-)fundamental representation of the gauge group $SU(N_c)$. We
choose $N_c<N_f<3N_c/2$, where the theory is in the free magnetic phase, and we take $N_{f0}$ of the flavors to
be massless, such that $0<N_{f0}<N_c$. This implies that, non-perturbatively, far away from the origin, the
theory has a runaway potential for the mesons associated with massless quarks (see, for instance, the reviews
\cite{Intriligator:1995au,Terning:2003th}). The other $N_f-N_{f0}$ flavors are massive but much lighter than the
strong coupling scale.

One can analyze this theory in the IR by using the Seiberg duality
\cite{Seiberg:1994pq}, which transforms the model above to an $SU(N\equiv N_f-N_c)$ gauge theory and matter
content of a gauge neutral $N_{f}\times N_f$ meson matrix $\Phi_i^j$ and $N_f$ flavors of (anti-)fundamental
dual quarks $q^i$ ($\tilde q_j$). The superpotential is
\begin{equation}\label{SQCD}
    \mathcal W=hTr^{\prime}(q^i\Phi_i^j\tilde q_j)-h\mu^2Tr(\Phi_{11}+\Phi_{22})+\text{non-perturbative},
\end{equation}
where $Tr'$ is taken over the $N$ color indices, $Tr$ is over flavor indices, and we parameterize
\begin{gather}\label{fields}
\Phi=\begin{pmatrix}(\Phi_{11})_{N\times N} & \Phi_{12} & X\\
\Phi_{21} & (\Phi_{22})_{(N_c-N_{f0})\times(N_c-N_{f0})} & Y
\\ \tX & \tY & (Z)_{N_{f0}\times N_{f0}} \end{pmatrix},\cr\hspace{2em}
q^T=\begin{pmatrix}\frac1{\sqrt2}(\chi_++\chi_-)_{N\times N} \\
\frac1{\sqrt2}(\rho_++\rho_-)_{(N_c-N_{f0})\times N} \\ (\sigma)_{N_{f0}\times N}\end{pmatrix},
\hspace{2em} \cr\tq=\begin{pmatrix}\frac1{\sqrt2}(\chi_+-\chi_-)_{N\times N} \\
\frac1{\sqrt2}(\rho_+-\rho_-)_{(N_c-N_{f0})\times N}  \\
(\ts)_{N_{f0}\times N}\end{pmatrix}.\end{gather}
Note that the model in the previous subsection amounts to the case $N_{f0}=N_c-N_{f0}=N_f-N_c=1$.

Again, rank conditions force us to expand around a SUSY breaking vacuum, as in (\ref{classical solution}).
Indeed, considering the F-terms for $\Phi_i^j$, the rank from the cubic superpotential coupling is at most
$N=N_f-N_c$ while the rank from the linear terms in the superpotential is $N_f-N_{f0}$. As long as $N_{f0}<N_c$
we cannot balance these terms and SUSY is classically broken. Interestingly, this condition is also necessary
and sufficient for runaway behavior, which is induced by non-perturbative dynamics.

So, the system settles into a SUSY breaking solution of the equations of motion,
\begin{gather}\label{vacuum}q^T=\begin{pmatrix}\mu\mathbb{I}_{N\times N} \\
0 \\ 0\end{pmatrix}, \hspace{2em} \tq=\begin{pmatrix}\mu\mathbb{I}_{N\times N} \\0
 \\0
\end{pmatrix}.\end{gather}
Expanding around this solution we discover, not surprisingly, a plethora of massive and massless modes, very
similar to the toy model analyzed in the previous subsection. A notable field is, of course, the $Z$ matrix
which remains massless even after a one-loop calculation for the same reasons as in our simplified model. All
the other modes are either massive at tree-level or gain some positive mass squared at one-loop. An unimportant
technical difference from the toy model is that now $\Im\chi_{-}$,$\Re\chi_{-}$ are eaten by the supersymmetric
Higgs mechanism.\footnote{However, the trace part remains massless as long as baryon symmetry is ungauged. The
real part of the trace becomes massive via one-loop effects and the imaginary part of the trace is an exact
Goldstone boson.}

Thus, again, we need to understand the dynamics of the $Z$ field near the origin. The global symmetry is
$SU(N_f-N_{f0})\times U(N_{f0})_\sigma\times U(N_{f0})_{\ts}\times U(1)_{\chi\rho} \times U(1)_R$ and is
spontaneously broken in the state~(\ref{vacuum}) to $SU(N_c-N_{f0})\times U(N_{f0})_\sigma\times
U(N_{f0})_{\ts}\times U(1)' \times U(1)_R$. The gauge symmetry $SU(N_f-N_c)$ is completely Higgsed. We can use
the subgroup $U(N_{f0})_\sigma\times U(N_{f0})_{\ts}$ to diagonalize $Z$ and make the eignevalues real. This
simplifies the mass matrix along pseudo-moduli space considerably. It is actually just several copies of the one
we considered in the previous subsection. In particular, these symmetry considerations imply that the quadratic
term takes the form $V^{(2)}\sim Tr(Z^\dagger Z)$, so to determine its coefficient it is enough to turn on a
single eigenvalue which is what we do in the following.

Let us first consider the non-gauge interactions. {}For the purpose of the two-loop computation, it is
straightforward to see that this model breaks up into $N(N_c-N_{f0})$ copies of the basic interactions we
considered in the simplified model. A straightforward way to see that is to reconsider any of the diagrams
depicted in Fig.~\ref{fig1}, e.g. the second diagram. There are $(N_c-N_{f0})$ possible $Y$ mesons (since only
one eigenvalue of $Z$ is turned on) and the color of the squarks has to be matched and summed over, so we get
another factor of $N$. Similar counting applies to the other two diagrams.

Now we have to turn on gauge interactions. The basic observation here is that the spectrum of vector multiplets
is supersymmetric over the whole moduli space~\cite{Intriligator:2006dd}. The reason is that gauge
symmetry is broken in the sector of $\chi$ which is decoupled in the mass matrix from SUSY breaking. Thus, our
criteria that there has to be a $\rho_\pm$ scalar and either $A$ or $B$ particles are still applicable. In these
circumstances, since gauge interactions are flavor diagonal, there are no vertices containing, for instance,
$A_\mu,\rho,\sigma$. Hence, there are no diagrams with particles from vector multiplets which contribute to
powers of $Z$ in the effective potential. One could worry about new interactions between quarks from
D-terms,
$$V_D=\frac{g^2}2\sum_A\left(Tr\, q^\dagger T_Aq-Tr\, \tq T_A\tq^\dagger\right)^2,$$
where the trace is over flavor indices. In the mass basis one can see that all the interactions $R_1^2R_2^2$,
where $R_1$ and $R_2$ are real scalars, cancel. So, there are no relevant contributions either from gauge
interactions or from D-terms (Intuitively, we do not expect non-trivial effects from D-terms in the absence of
accompanying fermionic loops.).

We conclude that this gauge theory exhibits instability near the origin, with no nearby minimum, plausibly
sloping to the runaway at large values of the $Z$ meson. The effective potential in the $Z$ direction takes the
explicit form
\begin{equation}\label{finalresult}
    V=h^6\mu^2\frac{N(N_{c}-N_{f0})}{(16\pi^2)^2}\left(-1-\frac{\pi^2}{6}+\ln4\right)Tr(Z^\dagger Z)+{\cal{O}}((Z^\dagger Z)^2).
\end{equation}

%%%%%%%%%%%%%%%%%%%%%%%%%%%%%%%%%%%%%%%%%%%%%%%%%%%%%%%%%%%%%%%%%%%%%%%%%%%%%%%%%%%%%%%%%%%%%%%%%%%%%%%%%%%%
\section{Brane Embedding}\label{braneembedding}

We now embed the gauge theory of subsection~\ref{full model} on intersecting branes in the type IIA string
theory. In subsection~\ref{bconfig} we present the brane construction and review the mapping of its parameters
to gauge theory. In subsection~\ref{bdynamics} we describe the perturbative brane dynamics -- the {\it
classical} forces between the branes -- and its interplay with the perturbative {\it quantum} dynamics found in
gauge theory.

\subsection{Brane Configuration}\label{bconfig} 
\begin{figure}
\begin{center}
\epsfig{file=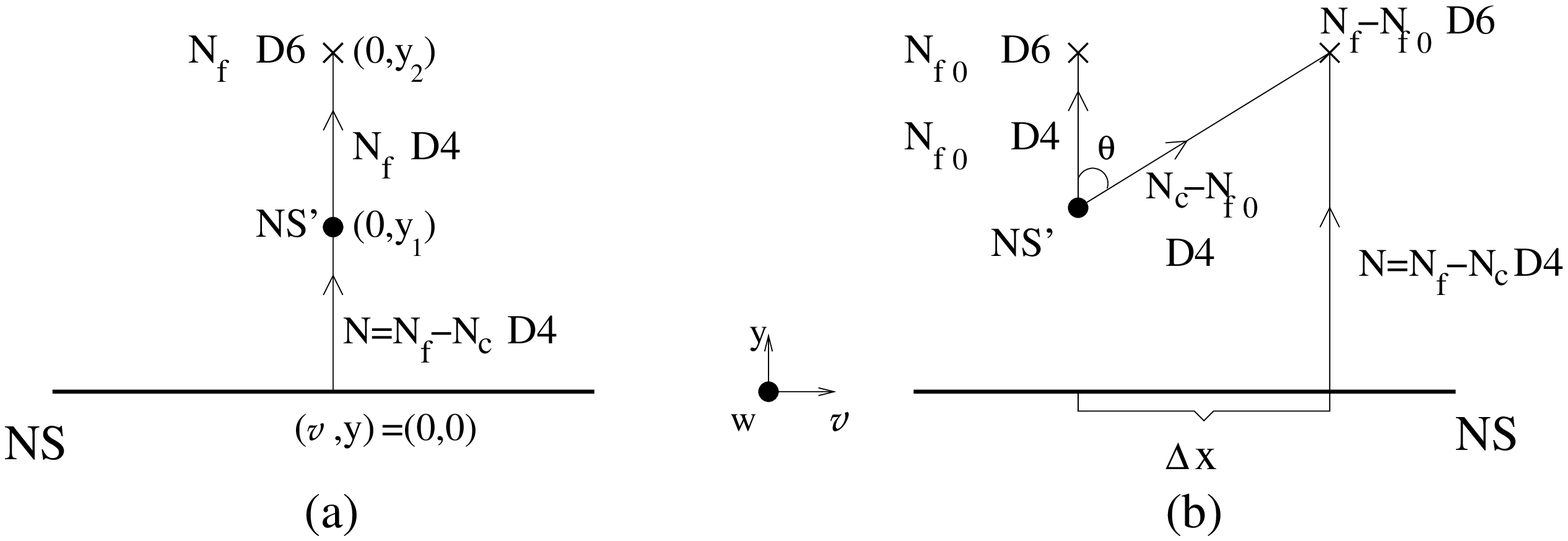,scale=0.5} \caption{(a) is the brane
construction of the magnetic theory with massless quarks.
(b) describes its deformation by a non-zero $\mu$ parameter.
In the electric langauge, $N_{f0}$ flavors are still
massless after this deformation.}\label{bconstruction}
\end{center}
\end{figure}

To construct the brane configurations in type IIA it is convenient to decompose
the $9+1$ dimensional spacetime as follows:
\begin{equation}
\mathbb{R}^{9,1}=\mathbb{R}^{3,1}\times
\mathbb{C}_v\times \mathbb{R}_y\times \mathbb{R}_{x^7} \times \mathbb{C}_w.
\end{equation}
The $\mathbb{R}^{3,1}$
is in the directions $(x^0,x^1,x^2,x^3)$,  common to all the branes. The complex planes $\mathbb{C}_v$,
$\mathbb{C}_w$ and the real line $\mathbb{R}_y$ correspond to
\begin{equation}
v=x^4+ix^5,\qquad w=x^8+ix^9,\qquad y=x^6.
\end{equation}
We begin with the brane configuration of  Fig.~\ref{bconstruction}(a), whose low energy limit is the magnetic
theory described in the previous section with $\mu=0$ \cite{Elitzur:1997fh} (for a review, see
\cite{Giveon:1998sr}).

Fig.~\ref{bconstruction}(a) presents a two dimensional slice $(x,y)$, where $x$ is a certain direction in $v$.
The line at the bottom of the figure stands for an NS5 brane, which is stretched in the direction $v$ and
located at $y=x^7=w=0$. We shall call it the NS brane. The bullet stands for another NS5 brane, which is
stretched in the direction $w$ and located at $v=x^7=0$ and $y=y_1>0$. We call it the NS' brane. The $\times$
denotes a stack of $N_f$ D6 branes, which are extended in the $(x^7,w)$ space and located at $v=0$ and $y=y^2$;
note that $y_2>y_1$. These are all the extended branes involved in our configurations.

We also have D4 branes which are stretched between extended branes. There are $N=N_f-N_c$ D4 branes stretched
between the NS and NS' branes, and $N_f$ D4 branes are stretched between the NS' and D6 branes. Arrows on the D4
branes indicate their orientation.

The low energy theory on the $N$  D4 branes stretched between the fivebranes is 3+1 dimensional $\mathcal{N}=1$
SYM with gauge group $U(N)$. Strings stretched between these $N$ ``color D4 branes'' and the $N_f$ ``flavor D4
branes'' correspond to $N_f$ fundamental chiral superfields $q^i,\tilde q_i$. Strings whose  both ends lie on
the flavor D4 branes give rise to gauge singlet superfields $\Phi_i^j$. These are coupled via the superpotential
\begin{equation}\label{bsup}
W_{\rm mag}=hq^i\Phi_i^j\tilde q_j.\end{equation} This magnetic theory is the Seiberg dual of $U(N_c)$ SQCD with
$N_f$ massless flavors \cite{Seiberg:1994pq}.

The mapping between the parameters of the brane construction and the gauge theory is the following. The
classical $U(N)$ gauge coupling $g_{\rm mag}$ is given by
\begin{equation}
g^2_{\rm mag}={g_sl_s\over y_1},
\end{equation}
where $g_s$ and  $l_s$ are the string coupling and length, respectively.
The Yukawa coupling
$h$ is given by
\begin{equation}\label{hhyy}
h^2={g_sl_s\over y_2-y_1}.
\end{equation}
Finally, the superpotential~(\ref{bsup}) has flat directions
corresponding to arbitrary expectation values of $\Phi$
while setting $q=\tilde
q=0$. In the brane picture, giving an expectation value $\langle \Phi_i^i\rangle$ corresponds to moving the
$i$'th flavor D4 brane to the location $w_i$ between the NS' and D6 branes.
A non-zero expectation value $\langle \Phi_i^i\rangle$ gives a mass $h\langle \Phi_i^i\rangle$ to the quarks
$q^i,\tilde q_i$. Geometrically, this corresponds to the length of a string stretched between the $i$'th flavor
brane and the color branes, and hence,
\begin{equation}
h\langle \Phi_i^i\rangle={w_i\over 2\pi l_s^2}~.
\end{equation}

Another deformation of this brane configuration, which is the main focus of this work, is to displace a stack of
$N_f-N_{f0}$ out of the $N_f$ D6 branes relative to the NS' brane in the $v$ direction. The resulting
configuration is shown in Fig.~\ref{bconstruction}(b). The separation  between the $N_f-N_{f0}$ D6 branes and
the NS' brane is denoted by $\Delta x$. This brane system is the one studied in
\cite{Ooguri:2006bg,Franco:2006ht,Bena:2006rg,Giveon:2007fk}; the latter focus on the special case $N_{f0}=0$,
while here we take $0<N_{f0}<N_c$. The configuration of Fig.~\ref{bconstruction}(b) is the energetically
favorable one.

After displacing the $N_f-N_{f0}$ D6 branes, only $N_{f0}$ of the flavor D4 branes, the ``massless flavor
branes,'' stay at their original position (with respect to NS').
On the other hand, $N\equiv N_f-N_c$ of the flavor branes
connect to the $N$ color branes and move with them to the position $v=\Delta x$, where they are stretched
between the NS and the $N_f-N_{f0}$ D6 branes in the $y$ direction. The remaining $N_c-N_{f0}$ flavor D4 branes
remain stretched between the NS' and the  $N_f-N_{f0}$ D6 branes, and hence are tilted in the $(y,v)$ space.

In the low energy gauge theory, this deformation amounts to adding to the magnetic theory (\ref{bsup}) a linear superpotential giving rise to the theory studied in the previous section, (\ref{SQCD}).
The mass parameter $\mu$ in the gauge
theory is related to $\Delta x$ by~\cite{Bena:2006rg}
\begin{equation}\label{stmu}\
\mu^2={\Delta x\over g_sl_s^3}.
\end{equation}
 Note that the brane pictures are reliable if the separations between the branes are
sufficiently large and the string coupling is small. We thus set $g_s\ll 1$ and $y_1,y_2-y_1>l_s$, but consider
the physics for generic values of $\Delta x$, similar to the study in \cite{Giveon:2007fk}. In the regime for
which $\Delta x>l_s$ perturbative string theory is reliable, and we can use it to study some aspects of the
brane dynamics. On the other hand, in the regime where $\Delta x$ is too small, the brane pictures are
misleading, since perturbative string theory is not reliable. In particular, for $\Delta x\ll g_sl_s$ we should
use gauge dynamics at low energies.

\subsection{Brane Dynamics}\label{bdynamics}
Several phenomena in gauge theory have simple analogs in the brane construction. The tilted branes break
supersymmetry. Furthermore, they can be displaced between the NS' and $N_f-N_{f0}$ D6 branes in the $w$
direction. This corresponds to the pseudo-moduli $\Phi_{22}$ in gauge theory.\footnote{More precisely, to the
expectation values $\langle(\Phi_{22})_i^i\rangle$; non-diagonal expectation values of $\Phi$ can be seen in the
brane pictures if one separates the D6 branes in the $y$ direction.} When $\Delta x$ is sufficiently large,
gravitational attraction of the tilted D4 branes to the NS brane fixes these moduli at $w=0$. In gauge theory,
an analogous effect is the stabilization of $\Phi_{22}$ at the origin by the one-loop effective potential.
Remarkably, it was observed \cite{Giveon:2007ef,Giveon:2007ew} that in a large class of brane constructions
gravitational attraction to the NS brane predicts phenomena which are realized in the low energy gauge theory
due to one-loop quantum effects.

The other pseudo-moduli in Fig.~\ref{bconstruction}(b) are the $N_{f0}$ deformations of the massless flavor
branes between the NS' and the $N_{f0}$ D6 branes. These correspond to the expectation values $\langle
Z_j^j\rangle$ in gauge theory. The location of the $N_{f0}$ D4 branes in $w$ is not fixed by an attraction to
the NS brane, since the NS and these D4 branes are mutually BPS. Indeed, the one-loop effective potential in
gauge theory does not fix the pseudo-moduli $Z$, as we have seen in the previous section.

We are thus led to consider subleading effects in the perturbative string theory regime (The analogous effect in
gauge theory is the two-loop effective potential we studied.). There are several effects here which play an
important role. Let us first focus on the NS' brane in Fig.~\ref{bconstruction}(b) and further concentrate on
the dynamics of the end-points of the fourbranes ending on it.\footnote{We thank David Kutasov for pointing out
the importance of the end-points dynamics, and for very helpful and interesting discussions. } These are
codimension-two objects in the world-volume theory of a type IIA fivebrane. To understand their interactions we
can consider the effective theory in three space-time dimensions. In this theory the end-points of D4 branes
correspond to localized sources giving rise to an electric field and some scalar fields, as
in~\cite{Callan:1997kz}.\footnote{ Recalling that the world-volume theory of a fivebrane in IIA string theory
does not contain vector fields it may be confusing that it appears (sourced by the end-point) after dimensional
reduction. The point is that the six-dimensional theory contains five scalars. Four are the usual Goldstone
modes which encode the shape of the fivebrane in ten dimension and the fifth is a compact scalar which has to do
with the M-theory circle. This compact scalar has a monodromy around the D4 end-point. Upon reducing to three
dimensions we can use Poincar\`e duality and turn this vortex source into a usual local electric source for an
Abelian three-dimensional gauge field. {}For a related analysis see~\cite{Aharony:1996xr}. We are grateful to Ofer
Aharony for very helpful and interesting discussions.}

More specifically, a single D4 brane ending on an NS' brane
at $w_0$ and going out in the direction $y$ gives rise, for large $|w-w_0|$, to the
following fields (in the normalization of~\cite{Giveon:1998sr}):
\begin{gather}y=g_sl_s\ln |w-w_0|,\hspace{1.5em} A_0={1\over l_s}\ln
|w-w_0|.\end{gather}
A fourbrane going out at an angle $\theta$ (like the tilted D4 branes in Fig.~\ref{bconstruction}(b))
from $w'_0$ has the following profile
(as follows by rotational invariance in the $xy$ plane):
\begin{gather}y=g_sl_s\cos\theta \ln |w-w'_0|,\hspace{1.5em} x=g_sl_s\sin\theta \ln |w-w'_0|,\hspace{1.5em}
 A_0={1\over l_s}\ln |w-w'_0|.\end{gather}
Since the world-volume theory of a single fivebrane is free we can use the superposition principle to construct
a solution for two such D4 branes,
\begin{gather}\label{electrostatics}y=g_sl_s(\ln |w-w_0|+\cos\theta \ln |w-w'_0|),\hspace{1.5em}
x=g_sl_s\sin\theta \ln |w-w'_0|,\cr
 A_0={1\over l_s}(\ln
|w-w_0|+\ln |w-w'_0|).\end{gather} It is straightforward to compute the binding energy of the system. Of
course, scalars of like charges attract while identical electric charges repel. When $\theta=0$ the system is
BPS and the forces conspire to cancel. A non-zero relative angle does not affect the electrostatic force, as is
evident in~(\ref{electrostatics}), but it decreases the attractive force from the exchange of $y$ bosons. There
is no overlap in the $x$ direction so there is no binding force from exchanges of $x$. Hence, the end-points
repel. The magnitude of the repelling force behaves, for large $|\Delta w|$, like
\begin{equation}\label{repulsion}
F(\Delta w)\simeq {g_s^2(1-\cos\theta)\over l_s|\Delta w|}~,\end{equation}
where $\Delta w=w_0-w'_0$.

Evidently, there are other forces acting in the system.
In general, separated non-parallel D branes in flat space
always attract since gravity dominates the RR repulsion.
So, far away from the NS' brane our $N_{f0}$ D4 branes
may feel some attraction.
However, the dominant effect near the NS' brane
is expected to be the Coulomb repulsion~(\ref{repulsion}).
To understand better the dynamics of this system one should solve the full
non-linear DBI action
(which should shed light on the short distance modifications of this Coulomb repulsion)
as well as analyzing better the closed string interactions involved.
Nevertheless, the considerations above strongly suggest
that the end-points repel each other and the origin at $w=0$ is
destabilized.

To recapitulate,
we presented an argument that the $N_{f0}$ D4 branes in Fig.~\ref{bconstruction}(b)
are destabilized in the brane dynamics regime,
nicely matching the field theory expectations.
The analysis in the perturbative brane regime is straightforward
and transparent compared to the intricate two-loop computation needed in the gauge theory.
However, the classical analysis above is not complete, but only presents some evidence
for what appears to be the correct dynamics.
It will be nice to perform more complete analysis of the various effects we
described above and to obtain quantitative predictions for the fate of this system
for generic separations of the brane.

\section{Comments on General Distributions of Masses}\label{comments}
There are some detailed implications of the results we obtained in the previous sections,
but here we restrict
ourselves to some qualitative features and postpone the complete phenomenological analysis to the future.
Consider massive SQCD with $N_f$ quarks in the free magnetic phase ordered as
\begin{equation}0<m_1\leq m_2...\leq m_{N_f},\end{equation}
where we take the mass matrix to be diagonal with positive real eigenvalues $m_i$.
We are interested in estimating how large should the hierarchy be, and among which
masses, such that the model is destabilized.

By the Seiberg duality we arrive at the theory of subsection~\ref{full model} with the superpotential
\begin{equation}\label{generic masses}
    \mathcal{W}=hq\Phi\tilde q-h\sum_{i=1}^{N_f}\mu_i^2\Phi_{ii},
\end{equation}
where $\mu_i^2=m_i\Lambda$ and $\Lambda$ is a strong coupling scale.
If $\mu_1=0$ then $\Phi_{11}$ is not lifted at one-loop, as we have seen.
Thus, the one-loop mass of $\Phi_{11}$ must be proportional to $\mu_1$. On the other hand, there is a
non-vanishing two-loop contribution. We know that it must be proportional to a combination of $\mu_1,..,\mu_{N_c}$
for the simple reason that if they all vanish the minimum is supersymmetric and the two-loop contribution
vanishes. The most dominant two-loop contribution comes from $\mu_{N_c}$.

We conclude that what is expected to affect the question of stability is primarily the ratio of $\mu_{N_c}$ and
$\mu_1$. The suppressing factor is, naively, ${4\pi\over h}$, the inverse loop expansion parameter. The correct
suppression factor is supposedly even smaller due to the loop coefficients we calculated.\footnote{For non-zero
$\mu_1$ we expect the two-loop result to contain dependence on the renormalization scale which can render a
precise estimate more complicated.}

\begin{figure}
\begin{center}
\epsfig{file=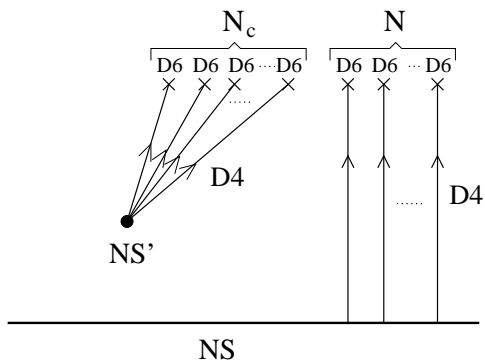,scale=0.5} \caption{The embedding of ISS with general masses into string theory. The
$N=N_f-N_c$ heavier flavors correspond to the vertical D4 branes, while the $N_c$ lighter flavors correspond to
the tilted D4 branes, one of whose ends lie on the NS' brane. As the mass of a light flavor is decreased, the D4
approaches a vertical line.} \label{distribution}\end{center}
\end{figure}

In the perturbative
brane dynamics regime a similar conclusion is made by very geometric and explicit means.
The theory with general masses in the magnetic description has a brane embedding shown in Fig.~\ref{distribution}. Now there are two competing forces: the attraction of the tilted
branes to the NS and the repulsion among them.
We shall call the tilted D4 brane corresponding to the $i$'th flavor the $\mu_i$ brane,
$i=1,\dots,N_c$.
The strength of the gravitational attraction of the $\mu_i$ brane to the NS is dictated by
$\mu_i$, as follows from eq. (\ref{stmu}).
Hence, the $\mu_1$ brane experiences the smallest attraction to the NS.
On the other hand, it is repelled from the other branes ending on the NS'.
The largest repulsion is due to its interaction with the $\mu_{N_c}$ brane,
since the angle between them is the largest.
It follows from eqs. (\ref{hhyy},\ref{stmu},\ref{repulsion})
that the strength of this repulsion is dictated by $h\mu_{N_c}$.
Thus, qualitatively, we see the same behavior as in the gauge theory: the stability of the
brane configuration is dictated by the ratio of $h\mu_{N_c}$ and $\mu_1$.

%%%%%%%%%%%%%%%%%%%%%%%%%%%%%%%%%%%%%%%%%%%%%%%%%%%%%%%%%%%%%%%%%%%%%%%%%%%%%%%%%%%%%%%%%%%%%%%%%%%%%%%%
\section{Summary}\label{summary}
We have analyzed several aspects of $SU(N_c)$ SQCD with $N_{f}-N_{f0}$ massive flavors and $N_{f0}$ massless
flavors in the range $0<N_{f0}<N_c<N_{f}<3N_{c}/2$. Through a two-loop computation (which was made feasible
using some simplifying observations and proper account of remnants of supersymmetric non-renormalization
theorems), we found that the field theory is in a runaway phase with no meta-stable states near the origin of
field space. We have also emphasized that our results may be important for model building inspired by ISS like
scenarios, since one is often forced to make some hierarchy of masses to take care of the longevity -- gaugino
masses tension (or to fix some other phenomenological problems, e.g. Landau poles). As we have shown, the
meta-stable minimum in massive SQCD with hierarchical masses may be destabilized due to two-loop radiative
corrections. It will be interesting to check what constraints are imposed by solving the above mentioned
phenomenological problems without inducing instability.

A similar picture was obtained for the brane embedding of this model, though by much more elementary means. This
provides an impressive test of the, yet mysterious, correspondence between the brane dynamics and gauge theory
in SUSY breaking configurations. The brane dynamics can be applied to other systems, leading to new non-trivial
``predictions'' in gauge theory. For example, for a general mass distribution, as in section~\ref{comments},
there are various brane predictions regarding the two-loop results in gauge theory, and it will be nice to test
them.

\begin{acknowledgments}
We thank O.~Aharony, M.~Berkooz, D.~Kutasov, Y.~Shadmi and D.~Shih for stimulating discussions. The work of A.~G
is supported in part by the BSF -- American-Israel Bi-National Science Foundation, by a center of excellence
supported by the Israel Science Foundation (grant number 1468/06), EU grant MRTN-CT-2004-512194, DIP grant H.52,
and the Einstein Center at the Hebrew University. The work of A.~K is supported in part by the Israel-U.S.
Binational Science Foundation (BSF) grant No. 2006071 and by Israel Science Foundation (ISF) under grant
1155/07. The work of Z.~K is supported in part by the Israel-U.S. Binational Science Foundation, by a center of
excellence supported by the Israel Science Foundation (grant number 1468/06), by a grant (DIP H52) of the German
Israel Project Cooperation, by the European network MRTN-CT-2004-512194, and by a grant from G.I.F., the
German-Israeli Foundation for Scientific Research and Development.
\end{acknowledgments}

%%%%%%%%%%%%%%%%%%%%%%%%%%%%%%%%%%%%%%%%%%%%%%%%%%%%%%%%%%%%%%%%%%%%%%%%%%%%%%%%%%%%%%%%%%%%%%%%%
\appendix
\section{Two-Loop Effective Potential}\label{append}
In this appendix we briefly review some of the results of the calculation of the two-loop effective
potential~\cite{Martin:2001vx}, which are used in this work. As we show in subsection~\ref{full model}, the
effects of gauging are irrelevant for our calculations in this paper. Hence, for simplicity we shall review here
interacting theories of scalars and fermions.

Consider a model with a set of \emph{real} scalars $R_i$ and Weyl fermions $\psi_I$. The masses of these
are given by
\begin{equation}
{\cal L}_{mass} = -\frac12 (m^2)^{ij}R_i R_j -\frac12 M^{IJ} \psi_I \psi_J +{\rm c.c.}~.
\end{equation}
We consider a basis where the mass-squared matrices $m^2_{ij}$ and
$M^2_{IJ}\equiv M^\dagger_{IK}M_{KJ}$ are already diagonal,
with eigenvalues $m_i^2$ and $m_I^2$, respectively.
Note that the (symmetric) fermionic matrix $M_{IJ}$ is
not necessarily diagonal.

The only possible renormalizable interactions in this theory are cubic and quartic interactions for the
scalars and Yukawa interactions of two fermions and a scalar. Following the conventions of~\cite{Martin:2001vx} we parameterize
them as follows:
\begin{equation}
{\cal L}_{int} = -\frac16 \lambda^{ijk}R_i R_j R_k -\frac{1}{24} {\lambda}^{ijkl}R_i R_j R_k R_l -
\left( \frac{1}{2} Y^{IJk} \psi_I \psi_J R_k + {\rm c.c.}\right)~.
\end{equation}
Note that the couplings $\lambda$ and $\lambda'$ are real and symmetric under the interchange of each
pair of indices. The Yukawa couplings $Y_{IJk}$ are symmetric under interchanges of spinor flavor indices $I$ and $J$.

\begin{figure}
\begin{center}
\epsfig{file=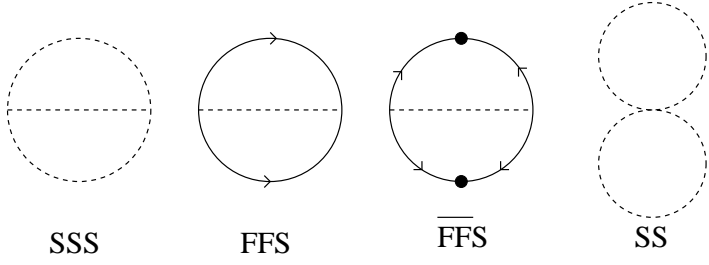,scale=0.5} \caption{Four possible diagrams which
contribute to the two-loop effective potential of a theory with interacting scalars
(represented by dashed lines) and fermions (solid lines).}\label{4dg}
\end{center}
\end{figure}

In the perturbative regime one can expand the effective potential as
\begin{equation}
V=V^{(0)}+\frac{1}{16\pi^2} V^{(1)} +\frac{1}{(16\pi^2)^2}V^{(2)} +\cdots ~.
\end{equation}
Generically, the two-loop potential $V^{(2)}$ depends on the renormalization scale, $Q$. The four possible
diagrams which can contribute to $V^{(2)}$ are depicted schematically in Fig.~\ref{4dg}. We will further refer
to these diagrams as SSS, FFS, $\overline{{\rm FF}}{\rm S}$ and SS respectively. Note that the diagram
$\overline{{\rm FF}}{\rm S}$ appears since the masses of fermions are not necessarily diagonal.

The contribution of each of these diagrams is parameterized by the following functions:
\begin{eqnarray}
V^{(2)}_{{\rm SSS}} & = & \frac{1}{12} (\lambda^{ijk})^2 f_{{\rm SSS}}(m_i^2, m_j^2, m_k^2)\\
V^{(2)}_{{\rm SS}} & = & \frac18 {\lambda}^{iijj} f_{{\rm SS}}(m_i^2, m_j^2)\\
V^{(2)}_{{\rm FFS}} & = & \frac12 |Y^{IJk}|^2 f_{{\rm FFS}}(m_I^2, m_J^2, m_k^2)\\
V^{(2)}_{\overline{{\rm FF}} {\rm S}} & = &
\frac14 Y^{IJk} Y^{I'J'k}M^*_{II'}M^*_{JJ'} f_{\bar{{\rm F}} \bar{{\rm F}} {\rm S}}(m_I^2, m_J^2, m_k^2)+{\rm c.c.}
\end{eqnarray}
The functions $f$ can be expressed in terms of three functions $I(x,y,z)$, $J(x,y)$ and $J(x)$ which are defined
as
\begin{eqnarray}
J(x) & = & x\left( \ln \frac{x}{Q^2} -1 \right)\\
J(x,y) & = & J(x)J(y)\\
I(x,y,z) & = & \frac12 (x-y-z)\ln \frac{y}{Q^2} \ln \frac{z}{Q^2} +\frac12 (y-x-z)\ln \frac{x}{Q^2}
\ln\frac{z}{Q^2} +\\ \nonumber &&\frac12 (z-x-y)\ln \frac{x}{Q^2} \ln \frac{y}{Q^2}+ 2x \ln \frac{x}{Q^2} +2y
\ln \frac{y}{Q^2} + \\ \nonumber &&2z \ln \frac{z}{Q^2} -\frac52 (x+y+z) - \frac12 \xi(x,y,z)~,
\end{eqnarray}
where $\xi$ is defined by
\begin{eqnarray}
\xi(x,y,z) &=& R \biggl( 2\ln \frac{z+x-y-R}{2z}\ln \frac{z+y-x-R}{2z} -
\ln \frac{x}{z}\ln\frac{y}{z}-\\ \nonumber
&& 2{\rm Li}_2 \frac{z+x-y-R}{2z} - 2{\rm Li}_2\frac{z+y-x-R}{2z} +\frac{\pi^2}{3}\biggl)~,
\end{eqnarray}
with
\begin{equation}
R = \sqrt{x^2+y^2+z^2 - 2xy -2xz -2yz}.
\end{equation}
In terms of $I$ and $J$, the functions $f$ are given by
\begin{eqnarray}
f_{{\rm SSS}}(x,y,z) & = & -I(x,y,z)\\
f_{{\rm SS}}(x,y) & = & J(x,y)\\
f_{{\rm FFS}}(x,y,z) & = & J(x,y) - J(x,z) -J(y,z) +(x+y-z) I(x,y,z)\\
f_{{\rm \overline{FF} S}}(x,y,z) & = & 2I(x,y,z)~.
\end{eqnarray}

In our specific model we need only $I$ functions with at least one argument vanishing,
so we give them explicitly
\begin{gather*}I(0,x,y)=(x-y)\left({\rm Li}_2(y/x)-\ln(x/y)\ln\frac{x-y}{Q^2}+\half (\ln\frac x{Q^2})^2-\frac{\pi^2}6\right) \cr -\frac52(x+y)+2x\ln\frac x{Q^2}+2y\ln\frac y{Q^2}-x\ln\frac x{Q^2}\ln\frac y{Q^2}.\end{gather*}
In the case that two arguments vanish it simplifies further
\begin{gather*}I(0,0,x)=-\half x(\ln\frac x{Q^2})^2+2x\ln\frac x{Q^2}-\frac52 x-\frac{\pi^2}{6} x.
\end{gather*}
However, the expression for $I(0,x,y)$ is still not very convenient since it contains terms which have no Taylor
expansion around the point $x=y$, which appears commonly in our expressions.
This is a spurious singularity which
cancels once all the terms are summed. To remove it once and for all we use Euler's identity for Dilogarithms
$${\rm Li}_2(x)+{\rm Li}_2(1-x)=-\ln x\ln (1-x)+\frac{\pi^2}6,$$ which gives
\begin{gather*}I(0,x,y)=(x-y)\left(
-{\rm Li}_2(1-y/x)-\ln (x/y)\ln\frac x{Q^2} +\half (\ln\frac x{Q^2})^2\right)\cr -\frac52(x+y)+2x\ln\frac
x{Q^2}+2y\ln\frac y{Q^2}-x\ln\frac x{Q^2}\ln\frac y{Q^2}.\end{gather*}

\end{document}